\documentclass[preprint]{elsarticle}

\usepackage[english]{babel}

\usepackage{amsmath}
\usepackage{amssymb}
\usepackage{cases}
\usepackage{natbib}
\usepackage{graphicx}
\usepackage[colorlinks=true, allcolors=blue]{hyperref}

\newcommand{\br}{{\bf \hat{r}}}
\newcommand{\btheta}{\mbox{\boldmath{$\hat{\theta}$}}}
\newcommand{\btau}{\mbox{\boldmath{$\tau$}}}

\newcommand{\MB}[1]{{\color{black} #1}}

\title{Linear instability in planar viscoelastic Taylor-Couette flow with and without explicit polymer diffusion}

\author[1,2]{Miguel Beneitez\corref{cor1}}
\ead{miguel.beneitez@manchester.ac.uk}

\author[3]{Soufiane Mrini}
\ead{soufiane.mrini@lisn.fr}

\author[2]{Rich R. Kerswell}
\ead{rrk26@cam.ac.uk}

\cortext[cor1]{Corresponding author}

\affiliation[1]{organization={Department of Mechanical and Aerospace Engineering}, addressline={The University of Manchester}, city={Manchester}, postcode={M13 9PT}, country={UK}}

\affiliation[2]{organization={DAMTP, Centre for Mathematical Sciences}, addressline={Wilberforce Road}, city={Cambridge}, postcode={CB3 0WA}, country={UK}}

\affiliation[3]{organization={Université Paris-Saclay, CNRS, Laboratoire Interdisciplinaire des Sciences du Numérique}, addressline={Orsay},postcode={91405}, city={Paris},country={France}}

\begin{document}

\begin{abstract}

Elastic turbulence has been found in computations of planar viscoelastic Taylor-Couette flow using the Oldroyd-B model, apparently generated by a linear instability (van Buel et al. {\em Europhys. Lett.}, {\bf 124}, 14001, 2018). We demonstrate that no such linear instability exists in the governing equations used unless some diffusion is added to the polymer conformation  tensor equation, as might occur through a diffusive numerical scheme. With this addition, the polymer diffusive instability (PDI) (Beneitez et al. ({\em Phys. Rev. Fluids}, {\bf 8}, L101901, 2023)) exists and leads to chaotic flows resembling those found by van Buel et al. (2018).
We show how finite volume or finite-difference discretisations of the governing equations can naturally introduce diffusive errors near boundaries which are sufficient to trigger PDI. This  suggests that PDI could well be important in numerical solutions of wall-bounded viscoelastic flows modelled using Oldroyd-B and FENE-P even with no polymer stress diffusion explicitly included.
\end{abstract}

\maketitle

\section{Introduction}

The addition of small amounts of polymer to a Newtonion solvent gives rise to a myriad of fascinating physical phenomena (including die swell, elastic recoil, and rod-climbing) \cite{datta2022perspectives,sanchez2022understanding}. It is particularly fascinating that chaotic dynamics -  so-called `elastic turbulence' - can be observed in the absence of inertia \cite{groisman2000elastic, steinberg2021elastic}. Elastic turbulence has practical applications in small scale flows, e.g. \MB{improved heat transfer efficiency \cite{traore2015efficient}} and efficient mixing where inertial turbulence cannot exist \cite{squires2005microfluidics}.  
Experimentally, elastic turbulence has been observed in rectilinear geometries \cite{pan2013nonlinear, qin2017characterizing, jha2021elastically,shnapp2022nonmodal} and geometries where the streamlines are curved  \cite{groisman2000elastic} with Taylor-Couette (TC) flow being the paradigm. 
Here the flow is contained between two concentric cylinders driven at different rotation rates and elastic linear instabilities due to hoop stresses are known to operate \cite{larson1990purely, shaqfeh1992effects, avgousti1993non, sureshkumar1994non, shaqfeh1996purely}. These hoop stress instabilities can be axisymmetric or non-axisymmetric but they all need structure in the axial direction, that is, the axial wavenumber has to be non-zero.  This means  that no hoop stress instability occurs in planar Taylor-Couette flow where the flow is restricted to be in the $(r,\theta)$-plane of cylindrical coordinates $(r, \theta,z)$ aligned with the axes of boundaries. Recent numerical work by \cite{van2018elastic}, however, reports the existence of a linear instability (see their figure 3)  and ultimately elastic turbulence in just such a limit. This raises the question as to what the linear instability mechanism is (e.g. see p57 in \cite{datta2022perspectives}). 

% now introduce rectilinear work....

About the same time as these numerical experiments, two new viscoelastic instabilities were found  for wall-bounded rectilinear flows within the Oldroyd-B and finitely-extensible-nonlinear-elastic-with-Peterlin-closure (FENE-P) models. The first, the `centre-mode instability',  was initially discovered in pipe flow \cite{garg2018viscoelastic}, then in channel flow \cite{khalid2021a, khalid2021b} but was confirmed absent from plane Couette flow \cite{KP24, Yadav24} (the instability was actually found a decade earlier in wall-less Kolmogorov flow at very different Reynolds numbers \cite{boffetta2005,LewyvKF}). 
A second linear instability - the `polymer diffusive instability' (PDI) - was found a little later in plane Couette flow \cite{beneitez2023polymer} and then channel and pipe flow \cite{couchman2023inertial}. PDI requires shear, a finite but non-zero polymer stress diffusion \cite{beneitez2023polymer} and a wall \cite{LewyvKF,lewy2024polymer} to operate. Neither instability has been discussed yet in the context of flows with curved streamlines such as the Taylor Couette set-up and the obvious
question is whether the linear instability found by  \cite{van2018elastic} is a manifestation of either of these two new instabilities. 

So motivated, we consider a FENE-P fluid with a large extensibility to remain close to the Oldroyd-B limit used in \cite{van2018elastic} and perform both linear stability analysis and direct numerical simulations of planar Taylor-Couette flow using a pseudo-spectral numerical codebase (based on the generic spectral solver Dedalus \cite{burns2020dedalus}). 
We demonstrate that in the absence of polymer diffusion, no linear instability exists  whereas in the presence of polymer diffusion, an instability in the form of a PDI mode is found. Our numerical simulations, which  are regularised via the explicit addition of polymer diffusion (a method which is frequently used in numerical simulations of viscoelastic flows \cite{alves2021numerical, dubief2023elasto}), show that this instability leads to non-trivial dynamics resembling those reported in \cite{van2018elastic}. To explain why these results rationalise  those in \cite{van2018elastic}, where no polymer diffusion is explicitly imposed there, we show that typical numerical discretisations (e.g. via finite differences) can introduce implicit diffusion particularly near boundaries and demonstrably excite PDI. 

The paper is structured as follows. We first introduce the formulation of the problem, and perform a linear stability analysis mirroring the configuration in \cite{van2018elastic}. We then discuss the linear stability results introducing explicit but small polymer diffusion into the problem. After this, we perform direct numerical simulations of the resulting linear instability, and then discuss how it is possible to have such an instability without explicit polymer diffusion. We finish the paper with the conclusions and a brief discussion. 

In review, we became aware of a parallel study \cite{Surya25} which focuses on three-dimensional perturbations and the possible interaction between PDI and the hoop-stress instability in Taylor-Couette flow. 

%
%   2
%
\section{Formulation and flow configuration}

We examine planar viscoelastic Taylor-Couette flow in which a polymer solution is confined between two concentric cylinders rotating at different rates and whose walls are separated by a distance $h:=r_o-r_i$ ($r_o/r_i$ are the outer/inner cylinder radii). The viscoelastic flow is modelled by the FENE-P equations 
\begin{align}
    \textit{Re}(\, \partial_t \mathbf{u}^* + (\mathbf{u}^*\cdot \nabla)\mathbf{u}^* \,) + \nabla p^* &= \beta\Delta \mathbf{u}^* + \label{eq:Ueq} (1-\beta)\nabla \cdot \mathbf{T}(\mathbf{C}^*), \\
    \partial_t \mathbf{C}^*+ (\mathbf{u}\cdot \nabla ) \mathbf{C}^* + \mathbf{T}(\mathbf{C}^*) &= \mathbf{C}^*\cdot \nabla \mathbf{u}^* +  (\nabla \mathbf{u}^*)^T\cdot \mathbf{C}^*+ \varepsilon\Delta \mathbf{C}^*,  \label{eq:Ceq}\\
    \nabla \cdot \mathbf{u}^* &= 0 \label{eq:divFree}
\end{align}
where
\begin{equation}
    \mathbf{T}(\mathbf{C}^*) := \frac{1}{\textit{Wi}}\left(f(\text{tr} \mathbf{C}^*)\mathbf{C}^*-\mathbf{I}\right), \qquad 
    f(s):=\left(1-\frac{s-3}{L_{\text{max}}^2}\right)^{-1}
\end{equation}
and $L_{\text{max}}$ is the maximum extensibility of the polymer chains ($L_{\text{max}} \rightarrow \infty$ recovers the Oldroyd-B model). $p^*$ denotes the pressure and $\mathbf{C}^*$ denotes the positive-definite conformation tensor which is the ensemble average of the tensor product of the end-to-end vector of the polymer molecules with itself. The equations have been non-dimensionalised by the distance between the walls, $h$,  and the speed of the outer cylinder, $U_o=r_o\Omega$, where $\Omega$ is the angular velocity and the inner cylinder is at rest. The parameter $\eta:=r_i/r_o$ quantifies the flow curvature. The Reynolds, \textit{Re}, and Weissenberg, \textit{Wi}, numbers are defined as
\begin{equation}
    \textit{Re}:=\frac{hU_o}{\nu} \quad \text{and} \quad \textit{Wi}:=\frac{\tau U_o}{h}
\end{equation}
where $\tau$ denotes the polymer relaxation time. The parameter $\beta:=\nu_s/(\nu_s + \nu_p)$ denotes the viscosity ratio, with $\nu_s$ and $\nu_p$ the solvent and polymer contributions respectively to the total kinematic viscosity, $\nu= \nu_s + \nu_p$. 
Equation ~\eqref{eq:Ceq} has a polymer diffusion term with diffusion coefficient \begin{equation}
\varepsilon:=\frac{1}{\textit{Re}\textit{Sc}}
\end{equation}
where the Schmidt number $\textit{Sc} = O(10^{6})$ at $Re=O(10^3)$ for a realistic polymer solution \cite{dubief2023elasto}. In numerical simulations, $\varepsilon$ is taken much larger - so $Sc \lesssim O(10^3)$ - and the diffusion term then considered more of a regulariser to maintain a positive-definite conformation tensor  and  numerical stability \citep{sid2018two,page2020exact}. When polymer diffusion is non-zero, boundary conditions are required on the polymer conformation equation, and we take the commonly-used  boundary condition that $\varepsilon=0$ at the walls \cite{sureshkumar97, dubief2020first, morozov2022coherent} (and for a general discussion \cite{LinK24}).  

As in \cite{van2018elastic}, we restrict the flow dynamics to the plane perpendicular to the axis of the cylinders.
\begin{equation}
\mathbf{u}^* = u_{r}^*(r, \theta,t) \br + u_{\theta}^* (r,\theta,t) \btheta.
\label{planar}
\end{equation}
The restricted velocity vector is 2-dimensional in the sense that there is no dependence on $z$ {\em and} there is no axial velocity component ($u_z^*=0$).
%
%   3
%
\section{Governing equations}
\subsection{The base flow}

We begin by examining the 1D azimuthal base flow in planar Taylor-Couette flow with explicit polymer diffusion. The governing equations for the base azimuthal flow  $U_\theta(r))$ and base polymer conformation tensor non-zero components ($C^b_{rr}(r), C^b_{r \theta}(r), C^b_{\theta \theta}(r), C^b_{rz}$ and  $C^b_{zz}(r)$) are
\begin{subequations}
    \begin{align}
        0 = \beta\left(-\frac{U_{\theta}}{r^2} + \frac{\partial_r U_\theta}{r}+\partial_{rr}U_{\theta}\right)       
        +\frac{1-\beta}{\textit{Wi}} &\left(
        \frac{f(\text{tr}({\bf{C^b}}))^2}{L^2}C_{r\theta}^{b}\cdot \text{tr}(\partial_r {\bf{C^b}}) \right. \nonumber \\
         &\left. \,\, +f(\text{tr}({\bf{C^b}}))
        \left[2\frac{C^b_{r\theta}}{r}+\partial_r C^{b}_{r\theta} \right]
        \right), \\
        \frac{1}{\textit{Wi}}\left(f(\text{tr}({\bf{C^b}}))C^b_{rr}-1\right) &= \varepsilon(\Delta {\bf C^b})_{rr}, \\
        2\frac{U_{\theta}}{r}C^b_{r\theta}+\frac{1}{\textit{Wi}}\left(f(\text{tr}({\bf{C^b}}))C^b_{\theta\theta}-1\right)&=2C^b_{r\theta}\partial_r U_{\theta}+\varepsilon(\Delta {\bf C^b})_{\theta\theta}, \\
        \frac{1}{\textit{Wi}}\left(f(\text{tr}({\bf{C^b}}))C^b_{zz}-1\right) &=\varepsilon(\Delta {\bf C^b})_{zz}, \\
        \frac{U_{\theta}}{r}C^b_{rr}+\frac{1}{\textit{Wi}}\left(f(\text{tr}({\bf{C^b}}))C^b_{r\theta}\right) &=C^b_{rr}\partial_r U_{\theta}+\varepsilon(\Delta {\bf C^b})_{r\theta}.
    \end{align}
\end{subequations}
Note that $C^b_{zz}$ is absent for Oldroyd-B but needs to be included for FENE-P due to the dependence of the stress-conformation tensor relation on $tr({\bf C})$. While this could be neglected in a two-dimensional version of the FENE-P model, it has minimal impact in the subsequent analysis and has been retained here.

\begin{figure}
    \centering
    \includegraphics[width=0.7\textwidth]{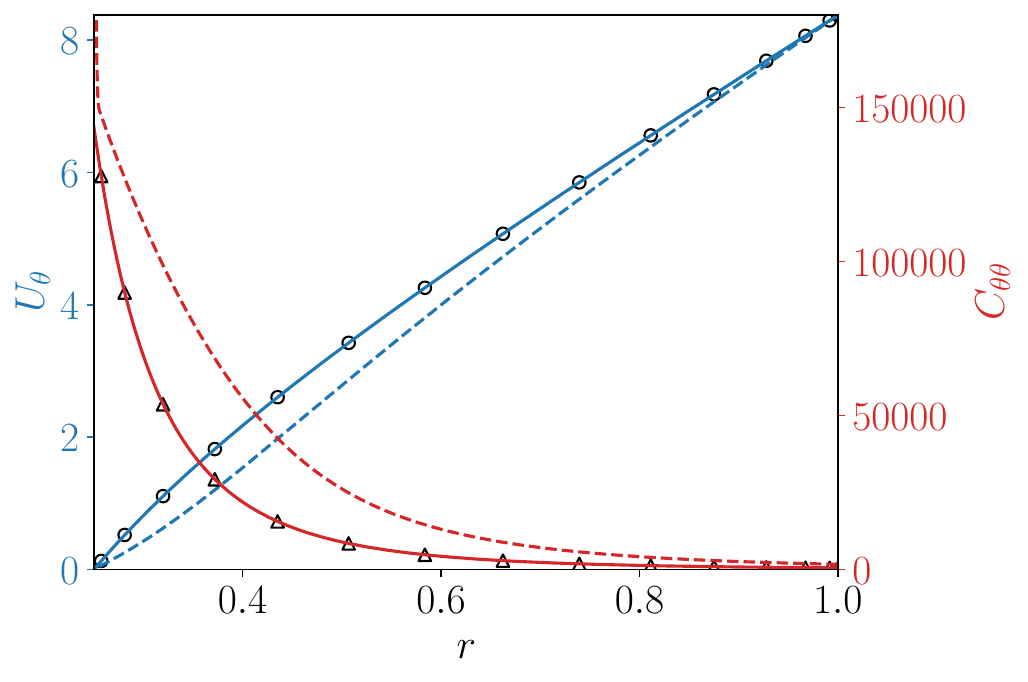}
    \caption{Base flow (in blue) in Taylor-Couette flow with parameters $\textit{Re}=0$, $\textit{Wi}=20$ and  $\beta=0.4$ consistent with those used in \cite{van2018elastic} (note their $\beta_{vB}=1.5$ translates into our $\beta=1/(\beta_{vB}+1)=0.4$ here).
    Resolution is $N=400$ Chebyshev polynomials.
    Black circles (triangles) indicate the analytical solution for $U_{\theta} \,(C_{\theta\theta})$ with $\varepsilon=0$. Solid blue (red) lines indicate the numerical solution for $U_{\theta}\, (C_{\theta\theta})$ using a scalar Laplacian for $\Delta \mathbf{C}$ with $\varepsilon=10^{-7}$. Dashed blue (red) lines indicate the numerical solution for $U_{\theta} \, (C_{\theta\theta})$ with a tensorial Laplacian for $\Delta \mathbf{C}$ with $\varepsilon=10^{-7}$.  This comparison shows that the scalar Laplacian provides a more appropriate  regularisation as the tensorial Laplacian affects the flow significantly even at such small $\varepsilon$.} 
    \label{fig:Fig1}
\end{figure}

%
% which Laplacian?
%
The natural choice for the Laplacian is the tensorial form, \MB{since its origin lies in the diffusivity of individual polymer molecules and therefore is proportional to the flux of the gradient of $\mathbf{C}$.} It is given by the various components
\begin{subequations}
\begin{align}
        (\Delta \mathbf{C})_{rr} &= \frac{\partial_{r} C_{rr}}{r} + \partial_{rr}C_{rr} + \frac{\partial_{\theta\theta}C_{rr}}{r^2}\boxed{+\frac{-2 C_{rr}-4 \partial_{\theta} C_{r\theta}+2 C_{\theta\theta}}{r^2}}, 
        \label{8a}\\
        (\Delta \mathbf{C})_{\theta\theta} & = \frac{\partial_{r} C_{\theta\theta}}{r} + \partial_{rr}C_{\theta\theta} + \frac{\partial_{\theta\theta}C_{\theta\theta}}{r^2}\boxed{+\frac{2 C_{rr}+4\partial_{\theta}C_{r\theta}-2C_{\theta\theta} }{r^2}},\\
         (\Delta \mathbf{C})_{zz} &=\frac{\partial_{r} C_{zz}}{r} + \partial_{rr}C_{zz} + \frac{\partial_{\theta\theta}C_{zz}}{r^2}, \\
        (\Delta \mathbf{C})_{r\theta} &=  \frac{\partial_{r} C_{r\theta}}{r} + \partial_{rr}C_{r\theta} + \frac{\partial_{\theta\theta}C_{r\theta}}{r^2}\boxed{+\frac{2\partial_{\theta}C_{rr}-4C_{r\theta}-2\partial_{\theta}C_{\theta\theta}}{r^2}}.  \label{8d}
\end{align}    
\end{subequations}
This maintains the coordinate-invariant property of  the equations and in particular correctly handles the singularity at the cylindrical axis if that is in the domain of interest. However, numerical errors will not generally respect this invariance property since they arise independently in each equation solved and so there is an argument that a scalar Laplacian on each component of the stress tensor could be a more appropriate model of any numerically-generated diffusion. This `scalar' choice would suppress all the boxed terms in the tensorial expressions (\ref{8a})-(\ref{8d}). It turns out that the scalar Laplacian also provides a better regularisation of the $\varepsilon=0$ situation compared to the tensorial Laplacian for the annulus under consideration (presumably because $\eta$ is quite small). Figure \ref{fig:Fig1} (a) shows the base flow for $\{\textit{Re}, \textit{Wi}, \beta, L_{\text{max}},\eta\}=\{0, 20, 0.4, 500,0.25\}$, and compares the analytical solution for $\varepsilon=0$ with the numerical base solutions for tensorial and scalar Laplacians at $\varepsilon=10^{-7}$. The figure demonstrates that the base flow is significantly modified by the tensorial Laplacian even at tiny $\varepsilon$ whereas it is not for the scalar Laplacian. Given this, we hereafter adopt the scalar version of the Laplacian acting on each component of the polymer conformation tensor to model the effect of implicit (numerical) diffusion.

%
% fig 1
%

\subsection{Linear stability analysis}

The linearised equations describing the evolution of infinitesimal perturbations $(u_r,u_\theta, c_{rr},c_{\theta \theta}, c_{zz}, c_{r \theta})$ read,
\begin{subequations}
    \begin{multline}
        \textit{Re}\left(\partial_t u_r + \frac{U_{\theta}}{r}\left( \partial_{\theta} u_r - 2u_{\theta} \right) \right) + \partial_r p  = \\
        \beta\left(\frac{\partial_r u_r}{r} + \partial_{rr}u_r + \frac{\partial_{\theta\theta} u_r}{r^2} - \frac{u_r}{r^{2}}-\frac{2\partial_{\theta}u_{\theta}}{r^2}  \right) + (1-\beta)\left(\frac{\partial_r \tau_{rr}}{r}+\frac{\partial_{\theta}\tau_{r\theta}}{r}-\frac{\tau_{\theta\theta}}{r}\right)
    \end{multline}
    \begin{multline}
        \textit{Re}\left(\partial_t u_{\theta} + u_r\partial_r U_{\theta} + \frac{U_{\theta}}{r}(\partial_{\theta} u_{\theta}+u_r) \right) + \frac{\partial_{\theta} p}{r} = \\
        \beta\left(\frac{\partial_r u_{\theta}}{r} + \partial_{rr}u_{\theta} + \frac{\partial_{\theta\theta} u_{\theta}}{r^2} - \frac{u_{\theta}
        }{r^{2}}+\frac{2\partial_{\theta}u_{r}}{r^2}  \right) + (1-\beta) \left(\partial_r \tau_{r\theta}+\frac{\partial_{\theta}\tau_{\theta\theta}}{r}+\frac{2\tau_{r\theta}}{r}\right),
    \end{multline}
    \begin{equation}
        \partial_t c_{rr}+u_r{\partial_r C^{b}_{rr} }+\frac{U_{\theta}\partial_{\theta}c_{rr}}{r} - 2\left(C^{b}_{rr}\partial_r u_r+ \frac{C_{r\theta}^{b}\partial_{\theta} u_{r}}{r}\right)+\tau_{rr} = \varepsilon(\Delta {c})_{rr},
    \end{equation}
    \begin{multline}
        \partial_t c_{\theta\theta}+u_r{\partial_r C^{b}_{\theta\theta} }+\frac{2u_{\theta}C_{r\theta}^{b}}{r} + \frac{U_{\theta}(\partial_\theta c_{\theta\theta}+2c_{r\theta})}{r} -\\
        2\left(C^{b}_{r\theta }\partial_r u_{\theta}+\frac{C_{\theta\theta}^{b}(\partial_{\theta}u_{\theta}+u_{r})}{r}\right) -2c_{r\theta }\partial_r U_{\theta} + \tau_{\theta\theta} = \varepsilon(\Delta {c})_{\theta\theta},
    \end{multline}
    \begin{equation}
        \partial_t c_{zz}+u_r{\partial_r C^{b}_{zz} }+\frac{U_{\theta}\partial_{\theta}c_{zz}}{r} + \tau_{zz} = \varepsilon(\Delta {c})_{zz},
    \end{equation}
    \begin{multline}
        \partial_t c_{r\theta} + u_r\partial_{r}C_{r\theta}^{b} + \frac{u_{\theta}C_{rr}^{b}}{r}+\frac{U_{\theta}\partial_{\theta}c_{r\theta}+c_{rr}}{r} - C_{rr}^{b}\partial_r u_{\theta} - \frac{C_{r\theta}^{b}(\partial_{\theta}u_{\theta}+u_r)}{r} - \\ C_{r\theta}^{b}\partial_{\theta}u_r-\frac{C_{\theta\theta}^b\partial_{\theta}u_r}{r} - c_{rr}\partial_r U_{\theta}+\tau_{r\theta} = \varepsilon(\Delta {c})_{r\theta},
    \end{multline}
    \label{eq:pertEqs}
\end{subequations}
where
\begin{subequations}
\begin{align}
    \tau_{ij} &= \frac{1}{Wi}\left(\text{tr}\ \mathbf{c}\frac{f^2(\text{tr}\ \mathbf{C}^{b})}{L_{\text{max}}^2}C^b_{ij}+f(\text{tr}\ \mathbf{C}^b) c_{ij}\right),\\
    \partial_{\theta} \tau_{ij} &=\frac{1}{Wi}\left(\partial_{\theta}\text{tr}\ \mathbf{c}\frac{f^2(\text{tr}\ \mathbf{C}^{b})}{L_{\text{max}}^2}C^b_{ij}+f(\text{tr}\ \mathbf{C}^b)\partial_{\theta} c_{ij}\right),\\
    \partial_r \tau_{ij}& =\frac{1}{Wi}\left(\partial_r\text{tr}\ \mathbf{c}\frac{f^2(\text{tr}\ \mathbf{C}^b)}{L_{\text{max}}^2}C^b_{ij} + \text{tr}\ \mathbf{C}^b\partial_r \left[\frac{f^2(\text{tr}\ \mathbf{C}^b)}{L_{\text{max}}^2}C^b_{ij}\right] + \right. \nonumber\\ 
    & \left. \hspace{4cm} c_{ij}\partial_r f(\text{tr}\ \mathbf{C}^b) +  f(\text{tr}\ \mathbf{C}^b)\partial_r c_{ij}\right).
\end{align}
\end{subequations}

Given the azimuthal symmetry of the base state, linear stability can be assessed by considered all infinitesimal perturbations of the form $\phi(r,\theta,t)=\hat{\phi}(r)e^{im(\theta-c t)}$, where $c=c_r+i c_i \in {\mathbb C}$.
By periodicity in $\theta$, $m \in {\mathbb Z}$  and the objective is to uncover situations where $c_i > 0$ which indicates exponential growth. The remaining eigenvalue problem for $c$ is a set of coupled ODEs in $r$. 

%
% 4   RESULTS
%
\section{Results}
\subsection{Polymer diffusive instability}
%
% Fig 2
% 
\begin{figure}
    \centering
    \includegraphics[width=0.49\textwidth]{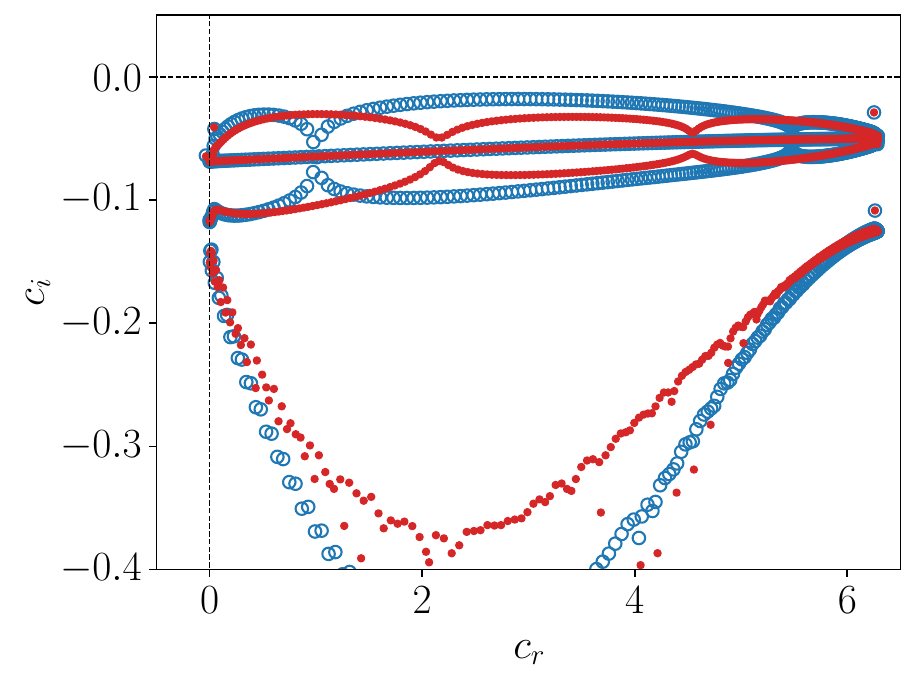}
    \includegraphics[width=0.475\textwidth]{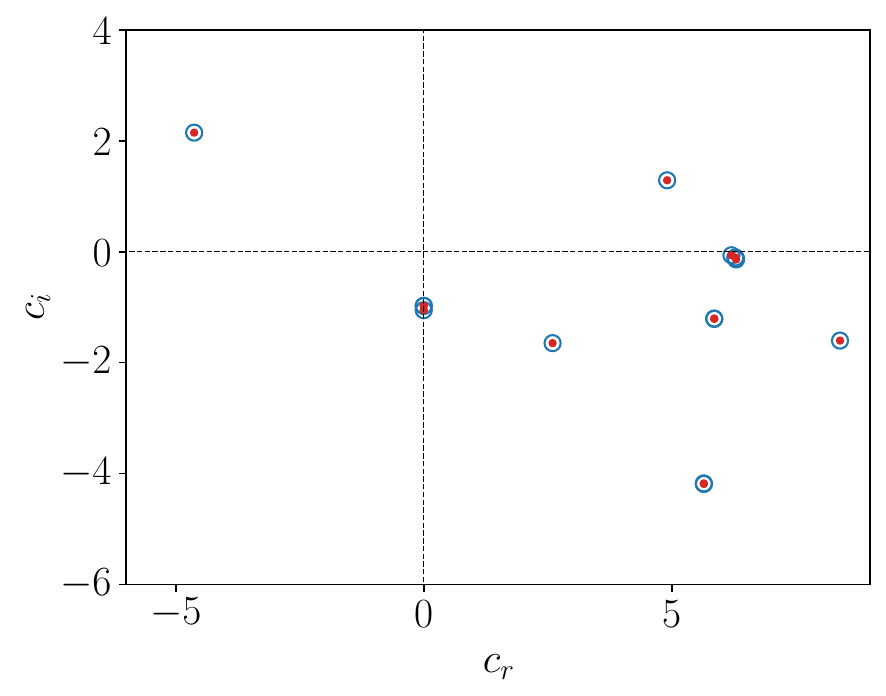}
    \caption{Left: Spectrum corresponding to the linear stability problem with parameters $\textit{Re}=0$, $\textit{Wi}=20$, $m=1$, $\beta=0.4$, $\eta=0.25$ and $\varepsilon=0$ and $L_{\text{max}}=500$ with resolution $N=300$ (blue dots) and $N=400$ (red dots). Right: Spectrum corresponding to the linear stability problem for the same parameters as left except now with $\varepsilon=0.1$. The leading eigenvalue $c =-4.628712+2.147339i$.}
    \label{fig:Fig2}
\end{figure}

We now present linear instability results using the spectral solver adapted from  \cite{beneitez2023polymer} for parameters $\{\textit{Re},\textit{Wi},\beta,L_{\text{max}},\eta\} = \{0,20,0.4,500,0.25\}$ as in Figure \ref{fig:Fig1} and for $m=1$. The maximum extensibility $L_{max}$ in our FENE-P model is set to a large value ($L_{\text{max}}=500$) so that our results can be directly linked to the Oldroyd-B results  of \cite{van2018elastic}. This choice of parameters is within the linear instability region reported in \cite{van2018elastic} (see their figure 3). The linear stability calculations are done with two different resolutions to ensure convergence of the results and to rule out the continuum spectrum. Figure \ref{fig:Fig2} (left) shows that in the absence of explicit polymer diffusion, $\varepsilon=0$, no eigenvalues cross the imaginary axis for $m=1$ or indeed any integer $m \in [-10,10]$ (not shown). These results indicate that the $\varepsilon=0$ base flow is  linearly stable which is not consistent with the results reported in figure 3 of \cite{van2018elastic}. Figure \ref{fig:Fig2} (right) shows the spectrum with the same parameters as used in Fig. \ref{fig:Fig2} (left)  but now taking $\varepsilon=10^{-1}$ for both the base flow and the perturbation. 
Introducing  polymer diffusion has a clear effect on the spectrum, which becomes discrete and two linearly unstable modes now exist. 

\begin{figure}
    \centering
    \includegraphics[width=0.5\textwidth]{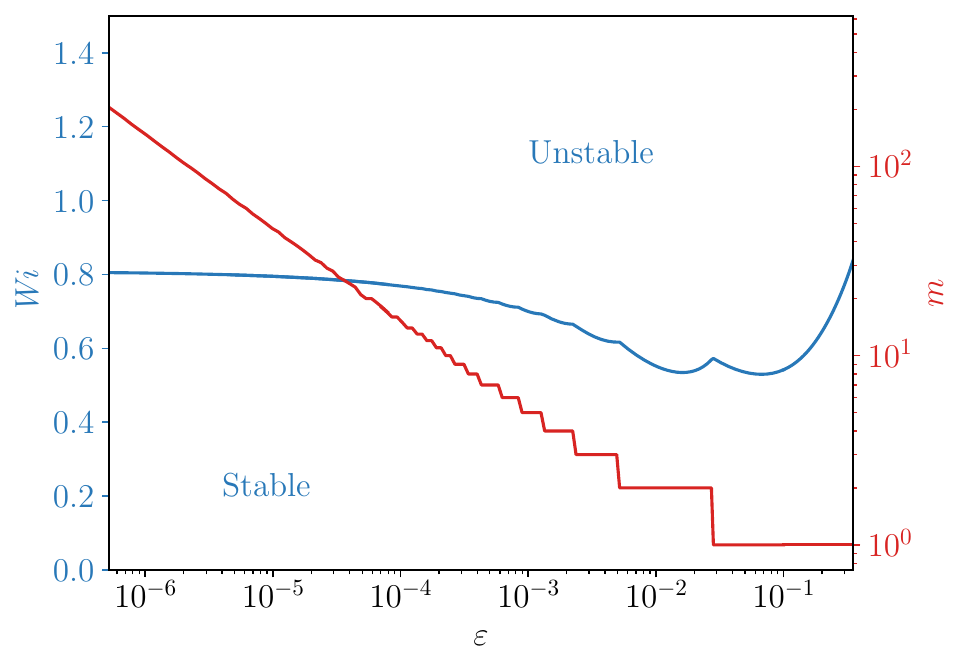}
    \includegraphics[width=0.49\textwidth]{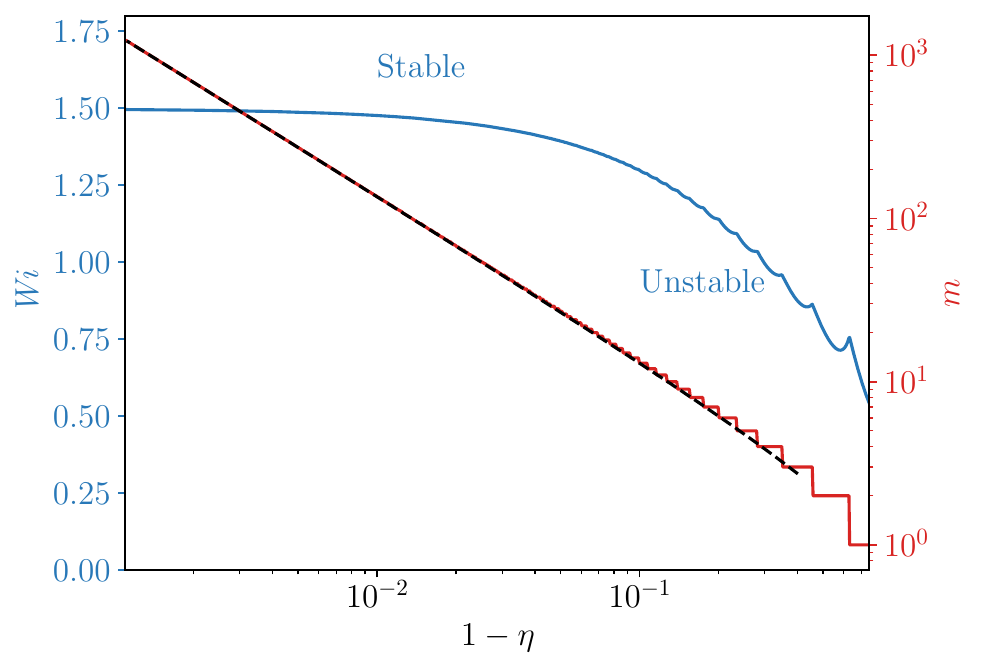}
    \caption{Left: Neutral stability curve in the $(\varepsilon, \textit{Wi})$ plane (blue) and corresponding wavenumber $m$ (red) for fixed parameters $\textit{Re}=0$, $\beta=0.4$, $L_{\text{max}}=500$ and $\eta=0.25$ using $N_r = 300$. 
    Right:  Neutral stability curve in the $(1-\eta, {\textit Wi})$ plane to characterise the effect of curvature. Fixed parameters are same as Fig. \ref{fig:Fig2} as left but for $\varepsilon=0.1$, instead of $\eta=0.25$. Black dashed line $(r_o+r_i)/\lambda_{crit,s}$, where $\lambda_{crit,s}$ denotes the critical wavelength in plane Couette flow. This confirms that the instability in the curved case is smoothly connected to the rectilinear limit.}  
    \label{fig:Fig3}
\end{figure}

The full neutral curve is shown in  Fig. \ref{fig:Fig3} (left). Three things are immediately clear as $\varepsilon$ decreases from $0.1$ towards $0$:
(i) the critical Weissenberg number, $Wi_c$, for instability tends to a constant as $\varepsilon \to 0$, 
(ii) $Wi_c$ does not vary much for finite values of $\varepsilon$, and 
(iii) the critical azimuthal wavelength goes to zero ($m \rightarrow \infty$). 
The particular value of $\varepsilon=0.1$ was chosen as the instability mode is then large scale: specifically, the most unstable mode has $m=1$ for $\varepsilon \gtrsim 3 \times 10^{-2}$. Fig. \ref{fig:Fig3} (right) illustrates how the linear instability identified here in planar Taylor-Couette flow can be continued in the geometry parameter $\eta=r_i/r_o$. Increasing curvature, i.e. increasing $1-\eta$, leads to a decreasing $Wi_c$ so making the flow more unstable. A black dashed line representing the critical wavenumber $(r_o+r_i)/\lambda_{crit,s}$  ($\lambda_{crit,s}$ denotes the critical wavelength in plane Couette flow) lies underneath the red curve corresponding to the current  instability indicating a connection with that in plane Couette flow \citep{beneitez2023polymer}. To further confirm this link, the instability scalings highlighted in  fig. \ref{fig:Fig3} (left) have the same PDI characteristics discussed in \cite{beneitez2023polymer, couchman2023inertial, lewy2024polymer}.

\subsection{Direct numerical simulations}
%
% fig 3
%

%
% fig 4
%
\begin{figure}
    \centering
    \begin{tabular}{cc}
    \
    \includegraphics[width=0.48\textwidth]{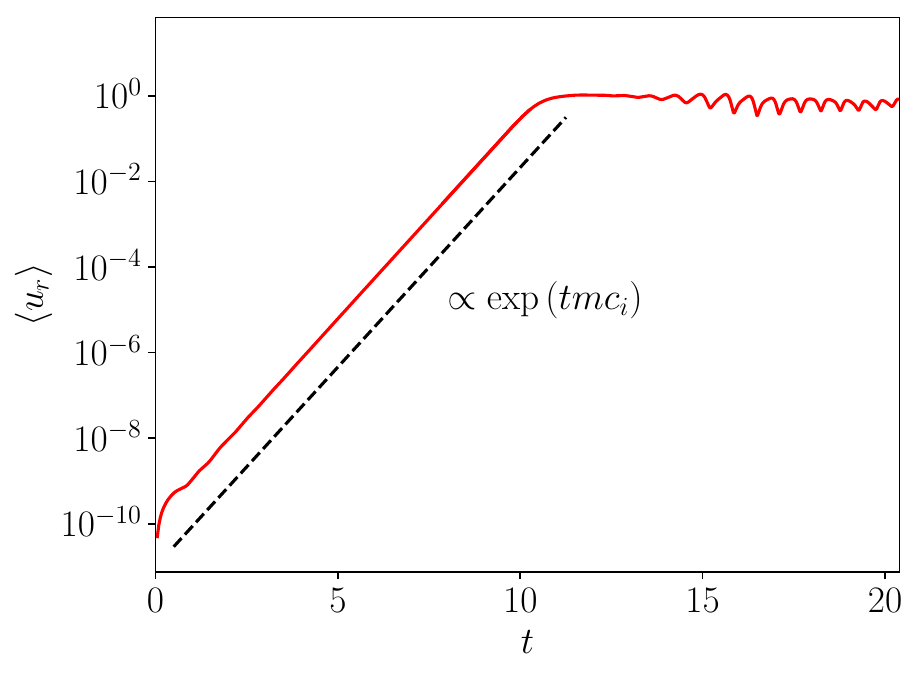} & \includegraphics[width=0.41\textwidth]{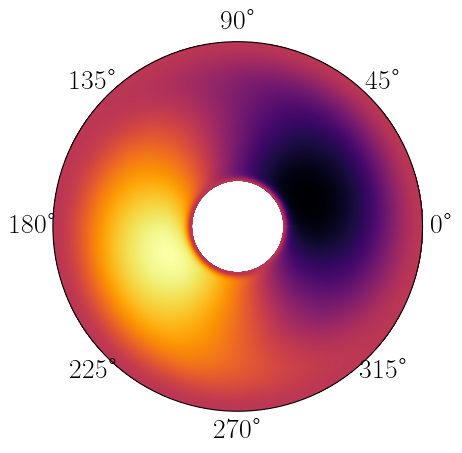}  \\
    \raisebox{-.5\height}{\includegraphics[width=0.51\textwidth]{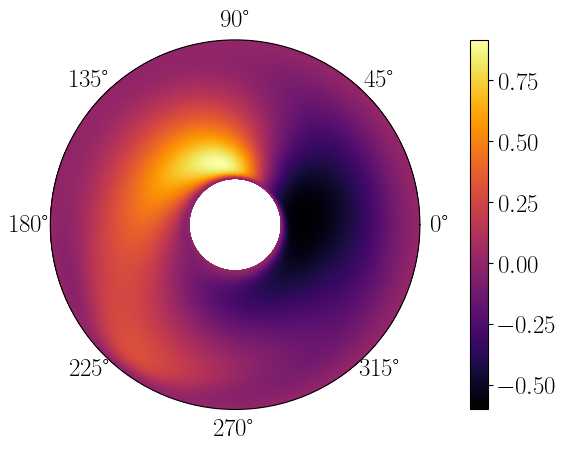}}
     & \raisebox{-.5\height}{\includegraphics[width=0.35\textwidth]{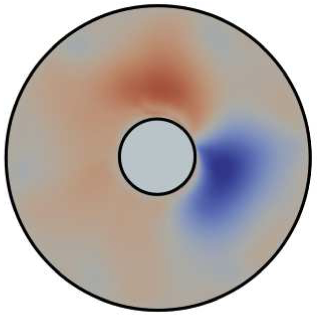}}
    \end{tabular} 
    \caption{Left top: time series of the volume average amplitude of the wall-normal velocity $\langle u_r \rangle$ for the DNS with parameters $\textit{Re}=0$, $\textit{Wi}=20$, $m=1$, $\beta=0.4$, $\varepsilon=0.1$ and $\eta=0.25$ with resolution $[N_r, N_{\theta}]=[256,256]$. The simulation is started with low amplitude $\mathcal{O}(10^{-11})$ noise and evolves into a chaotic state. Dashed black: theoretical growth predicted by the linear stability theory. 
    Right top: Snapshot of azimuthal velocity $u_r$ for the corresponding eigenmode.
    Bottom left:  the resulting chaotic state at $t=20.54$. 
    Bottom right: figure 2(a) from \cite{van2018elastic} showing the chaotic state found there at $Wi=12.6$.
    }
    \label{fig:Fig4}
\end{figure}

Having found that (scalar) polymer diffusion leads to a linear instability when none otherwise exists, we now check whether the nonlinear evolution of the (PDI) instability in Fig. \ref{fig:Fig2} is able to trigger elastic turbulence of the same kind as  seen in \cite{van2018elastic}. 
We use the Dedalus codebase \cite{burns2020dedalus} to solve the viscoelastic equations (\ref{eq:Ueq}-\ref{eq:divFree}) expanding the solutions using $[N_r,N_{\theta}]=[256,256]$  Chebyshev and Fourier modes in the radial and azimuthal directions respectively.  We choose the same set of parameters $\{\textit{Re},\textit{Wi},\beta,L_{\text{max}}, \eta,\varepsilon\}=\{0,20,0.4,500,0.25,0.1\}$ as above. The numerical simulations are initialised from the laminar state to which we add low amplitude noise $\mathcal{O}(10^{-11})$ to excite the linear instability. 
Figure \ref{fig:Fig4} (top left) shows that the time evolution for the observable $\langle u_r\rangle :=\int_V |u_r| dV$ displays exponential growth at the rate predicted by linear stability theory and Figure \ref{fig:Fig4} (top right) shows the radial velocity component for the corresponding eigenmode. The perturbation grows until nonlinear effects are significant (once $\langle u_r\rangle\sim 10^{-1}$), and then a chaotic (aperiodic) is observed for $t\geq 15$ in the current parameter set. These results demonstrate that explicit scalar diffusion in the polymer conformation tensor $\bf{C}$ can give rise to PDI which in turn is able to trigger chaotic dynamics. The chaotic state is characterised by a break-up in the radial symmetry of the radial velocity component as shown in Fig. \ref{fig:Fig4} (bottom left) and can be directly compared to figure 2(a) from \cite{van2018elastic} (reproduced here for clarity).

It is worth reiterating that in the absence of polymer diffusion we were unable to find any linear instability for the parameters and equations ostensibly solved in \cite{van2018elastic}. Moreover, the instability reported in \cite{van2018elastic} must have a growth rate of $\mathcal{O}(1)$ or so 
to grow initial noise of $\mathcal{O}(10^{-14})$ up to $\mathcal{O}(0.1)$ levels to trigger transition after $\sim 30$ time units (i.e. $10^{-14}\exp(1\times 30) \approx 0.1$). This approximate growth rate is in qualitative agreement with our results for PDI as shown in Fig. \ref{fig:Fig2}(right) where the growth rate is $\approx 2.15$. This  suggests that the numerical scheme (rheoTool) used in OpenFOAM for the results in \cite{van2018elastic} introduces implicit numerical diffusion in the simulation which leads to PDI and the chaotic dynamics associated with it.

\subsection{Implicit numerical diffusion}

We consider here a couple of examples of how diffusion can be introduced by numerical discretisation. Implicit numerical diffusion is difficult to exactly quantify as it depends on the solution (and so will vary spatially and temporally) and contains higher order contributions beyond second order derivatives. Any theoretical model taking a constant diffusion coefficient in front of a Laplacian is therefore limited. Nevertheless, we will show the emergence of similar phenomena to that observed for simple explicit diffusion in cases where implicit numerical diffusion is expected.

%
% rheolTool
%
\subsubsection{Finite volume}

In their work, van Buel et al. \cite{van2018elastic} use the rheoTool package which is a finite-volume solver that, for example,  has to extrapolate the value of stress tensor $\btau$ at the centre of a cell to that at the wall to estimate fluxes. This is done by using only the first order approximation (the first 2 terms) of the Taylor expansion
\begin{equation}
   \tau_{ij,w} = \tau_{ij,p} + D^1 \tau_{ij,w} \cdot \mathbf{d}_{pfw} + \frac{1}{2} D^2 \tau_{ij,w} \mathbf{d}_{pfw}^2 + h.o.t.
\end{equation}
where $\mathbf{d}_{pfw}$ is the vector connecting the centre of cell $p$ to the centre of the boundary face $w$ and 
\begin{equation}
D^{\alpha}\tau_{ij,p}=\frac{\partial^{\alpha_1+\dots +\alpha_n}\tau_{ij,p}}{\partial^{\alpha_1} x_1 \dots \partial^{\alpha_n} x_n}.
\end{equation}
This at once introduces a leading diffusive error term of the form $\varepsilon\Delta \tau_{ij}$ which will get multiplied by estimated gradients of the velocity field in the polymer equation. So a crude estimate for numerical diffusion coefficient for the stress is  $l^2|\nabla {\bf u}|$ where $l$ is the cell size. The presence of these terms provides a plausible explanation for the introduction of the PDI-like instability and subsequent chaotic dynamics described in the previous sections.

%
% fig 5
%
\begin{figure}
    \centering
    \includegraphics[width=0.49\textwidth]{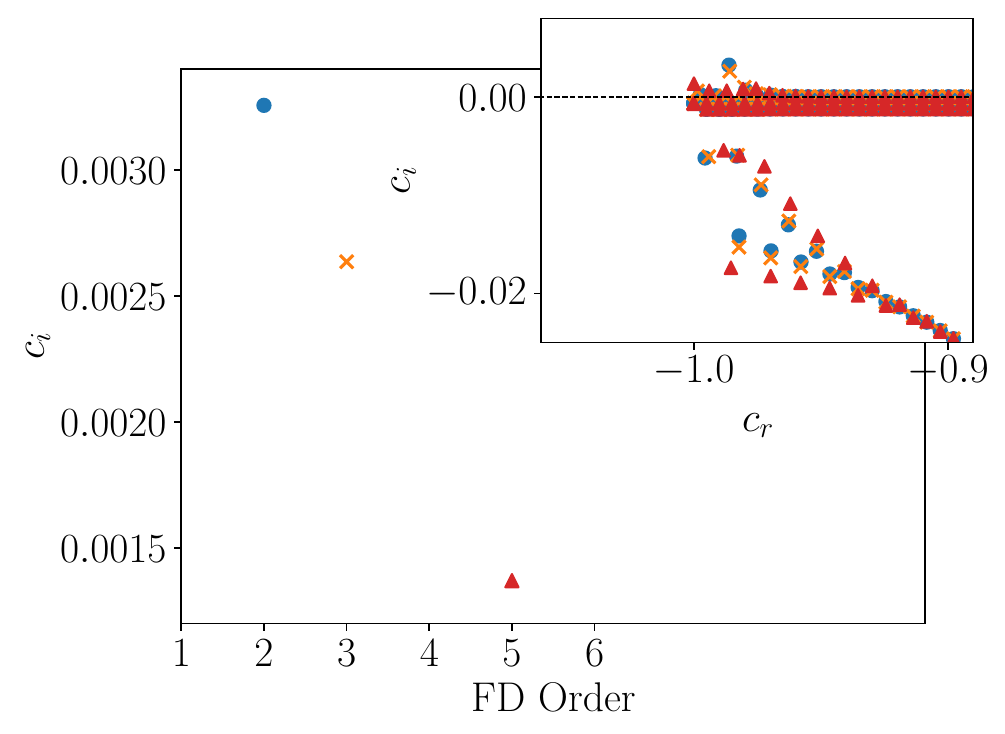}
    \includegraphics[width=0.47\textwidth]{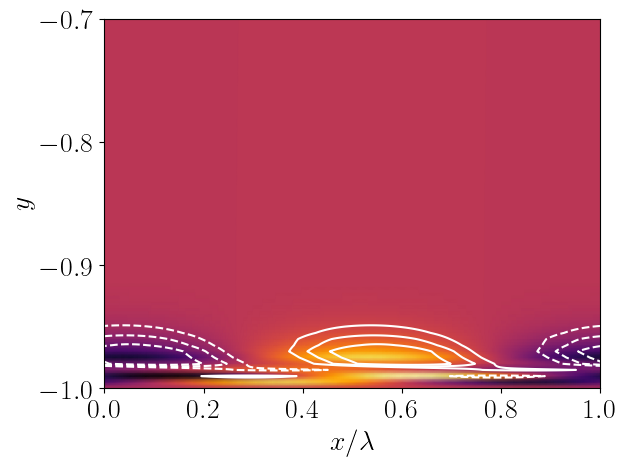}
    \caption{Left: Growth rates of the leading linear instability mode for the various order one-sided finite-differences for $\textit{Re}=0$, $\textit{Wi}=20$, $k=80$, $\beta=0.8$, $\eta=0.25$, $\varepsilon=0$ with resolution $N_y= 400$, schemes consider 2nd order (blue circles), 3rd order (orange crosses) and 5th order (red triangles).  Inset: full spectrum zoomed in the region close to $c_r=-1$. The figure shows that finite-difference schemes induce a linear instability.  Right: colour contours of $\text{tr}(\mathbf{C})$, streamline contours superimposed in white for positive (continuous lines) and negative (dashed lines)  for the leading eigenmode corresponding to the 2nd order one-sided finite-difference discretisation, i.e. leading blue eigenvalue on the left figure. One wavelength $\lambda=2\pi/k$, with $k=80$ is considered. The mode is a wall-mode akin to the PDI modes. Axes are not at scale.}
    \label{fig:Fig5}
\end{figure}

%
% FD
%
\subsubsection{Finite differences}

Section 4.1 and Fig. \ref{fig:Fig3} showed that the linear instability in \cite{van2018elastic} is a PDI mode which can be smoothly continued to the rectilinear flow configuration of 2D plane Couette flow and so we now solve the linear stability problem there with finite differences. Second-order centred finite differences can be used everywhere except at the boundaries.
This issue can be avoided if boundary conditions are available to replace imposing the relevant equation there (e.g. the velocity field and the momentum equation) but cannot be avoided for the polymer equation since there are no polymer boundary conditions when $\varepsilon=0$. In this case, the normal derivatives of the velocity  required to impose the polymer equation - for example, in the term $\mathbf{C}\cdot \nabla \mathbf{u} + (\nabla \mathbf{u})^T \cdot \mathbf{C}$ -  need to be estimated using one-sided derivatives. We do this using derivatives in turn of 2nd, 3rd and 5th order accuracy (see Tables 3.1-1 and 3.1-2 in \cite{Fornberg} reproduced in Table \ref{tab:table1} to explore their effect.

\begin{table}[]
\centering
\begin{tabular}{ccccccc}
                           & \multicolumn{6}{c}{Grid points (0) is at the wall} \\ \hline
\multicolumn{1}{c|}{Order} & 0       & 1      & 2     & 3     &  4  &  5 \\ \hline
\multicolumn{1}{c|}{2}     & -3/2    & 2      & -1/2  &       &      &  \\
\multicolumn{1}{c|}{3}     & -11/6   & 3      & -3/2  &  1/3  &      &  \\
\multicolumn{1}{c|}{5}     & -137/60 & 5      & -5    & 10/3  & -5/4 & 1/5  \\
\end{tabular}
\caption{Weights for the one-sided FD formulas used in Fig. \ref{fig:Fig5}
on an equi-spaced grid \cite{Fornberg}} \label{tab:table1}
\end{table}

Fig. \ref{fig:Fig5} (left, inset) shows part of the spectrum corresponding to the same parameter set used throughout the paper, $\{\textit{Re},\textit{Wi},\beta,L_{\text{max}},\varepsilon\}=\{0,20,0.4,500,0\}$ using $N_y=400$ discretisation points in the wall-normal direction and considering a streamwise wavenumber $k=80$. 
In the absence of explicit polymer diffusion ($\varepsilon=0$) where the flow is known to be linearly stable \cite{beneitez2023polymer}, a linearly unstable mode nevertheless appears caused by spatial discretisation errors. Consistent with this, the growth rate of this instability reduces as the order of the one-sided derivative is increased (main figure). This behaviour agrees with our understanding of the PDI, where the growth rate {\em at a fixed wavenumber} decreases as $\varepsilon$ becomes smaller \cite{beneitez2023polymer}. The structure of the linearly unstable mode when using a second-order one-sided  derivative at the walls is shown in Fig. \ref{fig:Fig5} (right).
The instability corresponds to a wall-mode which closely resembles the corresponding PDI mode in plane Couette flow: see fig 3(a) in \cite{beneitez2023polymer}.

\section{Conclusions}

In this paper, we have shown using a spectral solver that no linear instability is present in the  inertialess Oldroyd-B equations for planar Taylor-Couette flow ostensibly solved by van Buel et al. \cite{van2018elastic} despite their numerical simulations indicating one. Instead we suggest that the instability seen is due to the addition of implicit polymer diffusion by the numerical (finite-volume) scheme used and is, in fact, the polymer diffusive instability (PDI) found recently in rectilinear flows such as  plane Couette flow \cite{beneitez2023polymer}, channel and pipe flow  \cite{couchman2023inertial}. By explicitly adding polymer diffusion to the equations, we find a linear instability and ensuing chaotic dynamics  which resembles those found in \cite{van2018elastic}.

The correspondence cannot be exact as numerically-introduced diffusion  does not take the form of a simple diffusion term preceded by a constant coefficient added to the equations. Here we have argued that the better theoretical model of these errors is to add scalar diffusion to each polymer stress component instead of adopting a formal tensorial Laplacian on the basis that a) discretisation errors are not introduced in a coordinate-invariant way, and b) only the scalar Laplacian acts as a regularisation term (the tensorial Laplacian significantly alters the base flow due to the large curvature of the inner boundary). 

We have also tried to show how diffusive errors can get introduced into the equations by finite volume and finite difference discretizations. In the former case, which applies to the simulations in \cite{van2018elastic}, it is clear how extra diffusion in the stress field can get included by estimating the stress at cell walls from it values at cell centres. The situation for finite differences is less clear as one-sided derivatives at walls only introduce diffusion from the velocity field into the polymer equation rather than polymer stress diffusion. Nevertheless, this gives rise to a linear instability in the form of a wall mode akin to a PDI mode when there should be none. Reassuringly, the growth rate of this mode decreases with increasing order in the one-sided scheme used suggesting it will eventually stabilise at small enough grid size.

Our work could have significant implications for popular numerical codes, such as the widely used OpenFOAM\textsuperscript{\textcopyright} \cite{weller1998tensorial}, which might be susceptible to PDI-like instabilities owing to the discretisation used despite polymer diffusion not explicitly being included. For example, it is possible that the transition mechanisms to viscoelastic turbulence (ET and EIT) in other flows might rely on PDI-like modes, e.g. pipe flow \cite{lopez2019dynamics}.  This is potentially important as it is still unclear whether PDI is a physical phenomenon  or an artificial instability of the Oldroyd-B and FENE-P models (e.g \cite{Pandey25}). Clearly, there is an urgent need to appreciate how prevalent PDI is in other models and, more importantly, whether it, or indeed its secondary instabilities \cite{beneitez2024transition}, can be observed in experiments.

The authors gratefully acknowledge the support of EPSRC through Grant EP/V027247/1.

\bibliographystyle{unsrt}
\bibliography{main}

\end{document}